\documentstyle[12pt,fullpage]{article}
\newcommand{\fto}{\rightarrow}
\newcommand{\inr}{\in_{\mbox{R}}}

\newcommand{\Do}{{\bf do\ }}

\newcommand{\For}{{\bf for\ }}
\newcommand{\Endfor}{{\bf endfor}}

\newcounter{protoline}
\newlength{\boxwidth}

\newtheorem{proto_body}{Protocol}[section]
\newenvironment{protocol}[1]{
\setcounter{protoline}{0}
\begin{minipage}{\boxwidth}
\newcommand{\step}[1]
        {\addtocounter{protoline}{1} {\item [\ \ \ \
\arabic{protoline}:]{\small##1 } }}
\begin{proto_body}[ #1 ] \hspace*{\fill}\par
\begin{description} }{\end{description}\end{proto_body}\end{minipage}}

\newcommand{\proto}[2]{\ \\ \fbox{ \begin{protocol}{#1}#2
                                   \end{protocol}}\newline\medskip}

\newenvironment{protocol_cont}[0]{
\begin{minipage}{\boxwidth}
\addtocounter{proto_body}{-1}
\newcommand{\step}[1]
        {\addtocounter{protoline}{1} {\item [\ \ \ \
\arabic{protoline}:]{\small##1 } }}
\begin{proto_body}
\ \\
\begin{description} }{\end{description}\end{proto_body}\end{minipage}}

\newcommand{\indtheorem}{\begin{quote}\begin{theorem}}
\newcommand{\indtheoremend}{\end{theorem}\end{quote}}
\newcommand{\indlemma}{\begin{quote}\begin{lemma}}
\newcommand{\indlemmaend}{\end{lemma}\end{quote}}

\newcommand{\ket}[1]{| #1 \rangle}

\newcommand{\comment}[1]{}
\newcommand{\BB}{\{0,1\}}

\newcommand{\Sa}{\mbox{${\cal A}${\sl lice\/}}}
\newcommand{\Sap}{\mbox{${\cal A}${\sl lyson\/}}}

\newcommand{\Saa}{\mbox{${\cal A}${\sl lyson\/}}}
\newcommand{\Ra}{\mbox{${\cal B}${\sl ob\/}}}

\newcommand{\sumd}[2]{\displaystyle \sum\limits_{#1}^{#2}}

\newcommand{\pp}[2]{\mbox{pp.~#1\,--\,#2}}

\newcommand{\jofc}{{\it Journal of Cryptology}}

\newcommand{\crypto}[1]{{\it Advan\-ces in Cryptology ---
Proceedings of Crypto\,'#1}, \aou~19#1, Springer\,--\,Verlag}
\newcommand{\euroc}[2]{{\it Advanc\-es in Cryptology ---
Proceedings of Eurocrypt\,'#1}, #2~19#1, Springer\,--\,Verlag}
\newcommand{\focs}[2]{{\it Proceed\-ings of #1 \mbox{Annual} IEEE
Symposium
on the Foundations of Computer Science}, #2}

\newcommand{\aou}{\mbox{August}}

\newcommand{\nov}{\mbox{November}}
\newcommand{\dec}{\mbox{December}}

\begin{document}
\hyphenation{pre-sents pre-sent}
\title{Defeating classical bit commitments with a quantum computer}

\author{
Gilles Brassard\,\\
Universit\'e de Montr\'eal
\thanks{\,D\'epartement IRO, Universit\'e de Montr\'eal,
C.P. 6128, succursale centre-ville,\hfill\mbox{}
\mbox{Montr\'eal (Qu\'ebec), Canada H3C 3J7.  e-mail:
brassard@iro.umontreal.ca.}}
\and
Claude Cr\'epeau\,\\
McGill University
\thanks{\,School of Computer Science, McGill University,
Pavillon McConnell,3480 rue University,\hfill\mbox{}
\mbox{Montr\'eal (Qu\'ebec), Canada H3A 2A7.  e-mail:
crepeau@cs.mcgill.ca.}}
\and
Dominic Mayers\,\\
Princeton University
\thanks{Department of Computer Science, Princeton University,
\hfill\mbox{}
\mbox{Princeton, NJ 08544-2087, U.S.A..   e-mail:
mayers@cs.princeton.edu.}}
\and
Louis Salvail\,\\
BRICS
\thanks{BRICS, Basic Research in Computer Science of
        the Danish National Research Foundation,
      Department of Computer Science,University of \AA rhus,
Ny Munkegade, building 540, 
DK-8000 \AA rhus C, Denmark.\hfill\mbox{}\mbox{e-mail:
salvail@daimi.aau.dk}}
}

\maketitle

\abstract{It has been recently shown by Mayers that no bit commitment scheme
is secure if the participants have unlimited computational power and
technology.
However it was noticed that a secure protocol could be
obtained by forcing the cheater to perform a measurement.
Similar situations had been encountered previously
in the design of Quantum Oblivious Transfer.
The question is whether a classical
bit commitment
could be used for this specific purpose.
We demonstrate that, surprisingly, classical unconditionally concealing
bit commitments do not help.}

\section{Introduction} After Mayers obtained his general
impossibility theorem for bit commitment schemes
(see the Appendix of \cite{bcms97} and
\cite{dom:impos,mayers97}), different kind of ideas were proposed by
some of us with the hope to realize
unconditionally secure bit commitment. 
It was then
realized 
that these apparently promising ideas were also
ruled out by Mayers' attack.  These attempts contributed to enhance our
understanding of what is going on with quantum bit
commitment. 
However, no complete discussion on the
subject has ever been provided in the literature. 
The most interesting attempts were based on the use of a classical
bit commitment together with temporary assumptions on the
power of the cheater.
The idea was to use
the classical bit commitment to force the cheater
to perform a measurement.  This would be useful to realize
many quantum protocols other than quantum bit commitments.  Here our
objective is to bring out the general principles that
explain why this approach does not work in quantum cryptography.   

Before we proceed, let us briefly explain the notion of bit commitment
and its impact in quantum cryptography.  Quantum cryptography is often
associated with a cryptographic application called key
distribution~\cite{bb84,bbbss92} and it has achieved success in this
area~\cite{bc96}.  However, other applications of quantum mechanics to
cryptography have also been considered and bit commitment was at the
basis of most if not all of these other
applications~\cite{bc96,bbcs91,bcjl93,yao95}.  A~{\em bit
commitment\/} scheme allows \Sa\ to send something to \Ra\ that
commits her to a bit $b$ of her choice in such a way that \Ra\ cannot
tell what $b$ is, but such that \Sa\ can later prove to him what $b$
originally was.  You may think of this as \Sa\ sending a note with the
value $b$ written on it in a strong-box to \Ra\ and later revealing him
the combination to the safe.

The commitment obtained after the commit phase is
binding if \Sa\ cannot change the value of $b$ and it
is concealing if \Ra\  cannot obtain any information about $b$ without
the help of \Sa.  The commitment is secure if it is binding and
concealing.  The commitment is unconditionally secure if it is secure
against a cheater, either \Sa\ or \Ra, with unlimited technology and
computational power. In 1993 a protocol for quantum bit commitment,
henceforth referred to as~BCJL, was thought to be {\em provably
secure\/}~\cite{bcjl93}.  Because of quantum bit
commitment, the~future of quantum cryptography was very bright, with
new applications such as the identification protocol of Cr\'epeau and
Salvail~\cite{claude:louis} coming up regularly.

The trouble began in October 1995 when Mayers found a subtle flaw in
the BCJL protocol.  Though Mayers explained his discovery to many
researchers interested in quantum bit commitment \cite{BCJL:kaput},
his result was not made entirely public until after Lo and Chau
discovered independently a similar result in March 1996
\cite{lc96}. The result of Mayers was more general than the one
obtained by Lo and Chau, but both used the same basic idea.  The
result of Lo and Chau did not encompass the BCJL protocol in which \Ra\
can obtain an exponentially small amount of information. (In practice
a protocol is considered secure as long as \Ra\  cannot obtain more than
an exponentially small amount of information on the bit committed by
\Sa, that is, an amount of information that goes exponentially fast
to $0$ as the number of photons used in the protocol increases.)
However, the final version published by Lo and Chau \cite{lc96} used
the techniques previously used by Mayers \cite{BCJL:kaput} to prove
the non security of the BCJL protocol and any other protocol published
at the time.  So, the paper of Lo and Chau \cite{lc96} is a proper
account of these preliminary results.

\section{The general impossibility theorem}
Now, we review the general theorem \cite{mayers97}, which says that a 
quantum protocol that creates an
unconditionally secure bit commitment is simply impossible.  The main
additional difficulty in the general result is that it is easy to
think that measurements and classical communication could be used to
restrict the behaviour of the cheater during the commit phase, and thus
obtain a secure bit commitment.  In fact, after BCJL was shown not
secure, the spontaneous attitude was to try alternative quantum bit
commitment protocols by making some clever use of measurements and
classical communication~\cite{people}.  Some of these protocols were
proposed after Mayers obtained the general result in March 1996.  
All of these protocols were found not secure against Mayers' attack.

There exists two approaches to deal with measurements and classical
communication in quantum bit commitment protocols: an indirect
and a direct approach.  In the first proof of Mayers
(see \cite{dom:impos} and the Appendix of \cite{bcms97}) the indirect 
approach was used.  
It was shown that
any protocol in which classical information is used is equivalent to
another protocol in which no classical information is used.
Then it was shown that no such protocol is
unconditionally secure.  The advantage of this
approach is that, after the reduction is shown, the attack on the new
protocol is easy to describe and analyze because there is no classical
communication anymore.  The disadvantage is that we don't deal
directly with the issue of classical communication and measurements,
that is, the attack obtained against the new protocol is not the one
that applies on the original protocol.  The attack on the new protocol
does not include any classical communication, whereas in the original
protocol the cheater must communicate classically with the honest
participant (otherwise this honest participant will wonder what is
going on).

We emphasize that the proof of the reduction which is not that hard
must nevertheless explain why the cheater can still cheat in the
original protocol despite the fact that he is restricted by
measurements and decoherence which must occur because of classical
communication.  Otherwise the overall proof would simply miss the
important issue of classical communication -- it would not encompass
the protocols and ideas that have been proposed recently
\cite{bc96,claude,kent97a,kent97}.  Mayers preferred to use a more direct
approach without reduction in \cite{mayers97}.  So, Mayers' paper
\cite{mayers97} directly describes and analyzes the real attack that
must be performed by the cheater.

Lo and Chau also wrote a paper \cite{lc97} to discuss the issue of
quantum communication and other aspects of Mayers' result. They used
a variant of Yao's model for quantum communication.  The essence of
this model is that a third system is passed back and forth under the
control of each participant when it is their turn~\cite{yao95}.  Mayers'
attack works fine in this model, and it is indeed important to verify
that the attack works in such a reasonable model.  With regard to
classical communication, the discussion of Lo and Chau \cite{lc97} is
similar to the indirect approach of Mayers. 

Now, let us consider the attack.  Of course, we are interested in the
attack on the original protocol.  The attack on the new protocol is
just a construction in a proof.  We emphasize that in both approaches,
with a reduction or without a reduction, the attack on the original
protocol is the same.  Here we focus on the part of the attack that
must be performed during the commit phase.  (The remainder of the
attack, which is performed after the commit phase, is the same as when
there is no classical communication, so it creates no additional
difficulty.)  One ingredient in the attack is that the cheater keeps
everything at the quantum level except what must be announced
classically.  Assume that at some given stage of the commit phase, a
participant has normally generated a classical random variable $R$,
performed measurements to obtain an overall outcome $X$, and shared
some classical information $Y$ with the other participant as a result
of previous communication.  Now, assume that this participant is the
cheater and that the protocol says he must transmit some classical
information $f(R,X,Y)$, which for simplicity we assume is a binary
string.  One might think that the cheater must have generated the
random variable $R$ and the outcome $X$ in order to
be able to compute and send $f(R,X,Y)$.  However, the cheater does not
have to do that.  As we explain later, he can do the entire computation 
of $f(R,X,Y)$,
including the computation of $R$ and the measurements, at the quantum
level.  Only $Y$ needs to be classical.  Then he can measure the bits
of the string $f(X,R,Y)$ (only these bits) and send them to the other
participant. 
The final result is that all information is kept at the quantum
level, except what must be sent classically to the other
participant. As explained in \cite{dom:impos,mayers97} (see also the
Appendix of \cite{bcms97}) this strategy performed during the commit phase 
either allows
\Ra\  to obtain some information about the bit committed by \Sa,
without any help from \Sa, or else allows \Sa\  to change her mind
after the commit phase.  

To understand how the cheater can perform the same algorithm at
the quantum level,  it is useful to keep in mind that any
classical process can be seen as a special kind of quantum phenomena.
Therefore, in principle no modification is required because the
classical algorithm already describes a quantum process.  It is
sufficient to describe the exact same algorithm in the viewpoint
that every phenomena corresponds to a unitary transformation.
The standard way to do that is the following.  Let ${\cal C}(\omega_0)$
be the initial state of a measuring apparatus.  A measurement
on a state $\psi$ becomes a unitary transformation
that maps $\psi \otimes {\cal C}(\omega_0)$ into $\sum_i \psi_i \otimes 
{\cal C}(\omega_i)$ where
${\cal C}(\omega_i)$ is the state of the measuring apparatus associated
with the outcome $\omega_i$ and $\psi_i$ is the corresponding (unnormalized)
final state of the measured system.  The generation of a random
variable $r$ with probability $p(r)$ corresponds to the creation
of a superposition $\sum_r \sqrt{p(r)} {\cal C}(r)$  where
the states ${\cal C}(r)$ are orthonormal states that encode the
random classical information $r$.   In particular,  the
state $\psi = \alpha {\cal C}(0)
+ \beta {\cal C}(1)$ corresponds to a random bit that takes the values
$0$ and $1$ with probability $|\alpha|^2$ and $|\beta|^2$ respectively.
Now, suppose that a function $f(x)$ must be computed.  The output
of the function $f$ requires a new register denoted $F$.  Initially,
the registers $X$ and $F$ are in state 
$\sum_x \lambda_x {\cal C}(x) \otimes {\cal C}(0)$, that is, the value
$x$ for the input occurs with probability $|\lambda_x|^2$ and the
register $F$ is initialized at $0$. The result of
the computation of $f$ is the state $\sum_x \lambda_x {\cal C}(x) \otimes
{\cal C}(f(x))$.   
When we say that the cheater performs the same algorithm
at the quantum level, we mean that the classical states
${\cal C}(x)$, ${\cal C}(r)$, etc.\  are replaced by orthogonal quantum states
of truly quantum systems.  
These techniques will become clear when
examples are discussed in the next section.

Despite the fact  that formally the algorithms are identical, 
the cheater will have more flexibility later on if he performs
his algorithm at the quantum level rather than at the classical
level.  It is not true that these
two levels are equivalent.  For example, the truly quantum state
$\alpha |0\rangle + \beta |1\rangle$ can be unitarily mapped
into the state $|0\rangle$, but this is not true for the
corresponding classical state because a part of the overall states 
${\cal C}(0)$ and 
${\cal C}(1)$ is encoded in an irreversible manner in the environment 
or the classical
apparatus, etc.  Therefore, one would
like to find a way to force the cheater to perform real measurements,
as requested in the honest protocol.  This would be useful not
only to realize quantum bit commitment protocols, but to
realize many other quantum protocols, including the important
quantum oblivious transfer protocols~\cite{crepeau94,bbcs91}.   

A better
understanding of the situation came after Cr\'epeau proposed a 
quantum protocol \cite{bc96,claude}
that uses a computationally secure classical bit commitment
\cite{BCC88,NOVY} as a subprotocol.  The idea was to rely temporarily
on the limitation (in speed) on the cheater during the commit phase to
force him to perform some measurements.  The hope
was that this short-term assumption could be dropped after the commit
phase so as to obtain a quantum bit commitment not relying on any
long-term assumption.   Salvail also proposed a protocol in which two
participants, \Sa\ and \Saa\ say, want to commit a bit to \Ra.  \Sa\
and \Saa\ are sufficiently far apart that they cannot communicate
during the commit phase.  Again the hope was that this temporary
restriction on the cheaters during the commit phase would be
sufficient to obtain a secure quantum bit commitment not relying on
any long-term assumption.

However, after some thoughts, one realizes that the cheater in Mayers'
attack performs the honest algorithm: the only difference is that he
performs this honest algorithm at the quantum level.  Therefore, if
the cheater has the power to perform the honest protocol (which he
must have) and has the technology to store information at the quantum
level, then he has the power to cheat during the commit phase, despite
the fact that he has not the power to break the computationally secure
bit commitment efficiently, or despite the fact that \Sa\  and
\Saa\  cannot communicate during the commit phase.  After the
commit phase, the rule of the game is that we must drop the assumption
on the computational power of the cheater, so the fact that a
computationally secure bit commitment was used is irrelevant: the
proof applies.  

\section{Quantum attacks}
Here, we analyze the possibility
to use classical bit commitment protocol to force the
cheater to perform a measurement.    
Our conclusion is that, surprisingly, a whole class of
classical BC schemes (that are perfectly concealing) fail miserably
in this scenario. Our result is illustrated with the
computational BC scheme of Naor, Ostrovsky, Venkatesen and Yung \cite{NOVY},
and the two-prover BC scheme of Ben-Or, Goldwasser, Kilian and Wigderson 
\cite{BGKW}.
The basic idea can be used regardless of the BC scheme.

The attack is inspired from the discussion of the previous section, 
but we will focus on the fact that the objective (defeated 
by the attack) is to force a measurement rather that an entire
protocol.  In the following, the goal of the protocol is
to force  a measurement using a classical bit commitment. 
In the cheating protocol, \Sa\ creates a superposition of all
possible honest strategies.
For example, suppose that \Sa\ is given a quantum state
$\psi = \alpha |0\rangle + \beta |1\rangle$ to measure
and that Bob is expecting a commitment to the outcome.  
In the cheating protocol, Alice measures the state $\psi$ at the quantum 
level as explained in the previous section.  Then she performs the
commit part of the (classical) protocol at the quantum level as if she
committed to the outcome of the measurement.  At the end of the
commit phase, this outcome is entangled with other registers
on Alice's side, but it is still in superposition. At
this point there are two possibilities.
\begin{itemize}
\item[$\bullet$] if unveiling is requested, she measures her
remaining quantum state and successfully complete the protocol as if
she had been honest all the way!

\item[$\bullet$] if no unveiling is requested, she undoes the
entanglement in such a way as to recover the state $\psi$ that was
given to her to start with, completely untouched!
\end{itemize}
This completely defeats the purpose of the BC scheme if the
goal was to force a measurement.  
We now illustrate this principle with two examples.

\subsection{NOVY Bit Commitment Scheme}
In a computational scenario two techniques reduce bit commitment to
very general cryptographic assumptions: the protocol of Naor 
\cite{Naor}
reducing unconditionally binding and computationally concealing bit
commitment to pseudorandomness, and the protocol of
Naor, Ostrovsky, Venkatesen and Yung \cite{NOVY} reducing computationally
binding and unconditionally concealing bit commitment to one-way
permutations. We restrict our attention to this second result and
first present their construction.  In the following, $\pi:\BB^n \fto
\BB^n$ denotes a one-way permutation.

\proto{$NOVY/Commit(b)$}{
\step{\Sa\/ picks $x \inr\BB^n$, and computes $y := \pi(x)$,}

\step{\For $i\in\{1,...,n-1\}$ \Do}

\step{\Ra\/ picks a hash vector $h_i \inr\BB^n$ and announces it to
\Sa,}

\step{\Sa\/ announces $r_i := h_i \cdot y$ to \Ra,}

\step{\Endfor}

\step{Let $y_0,y_1$ be two solutions to
$\{ r_i = h_i \cdot y_* \}_{1\leq i< n}$
in some fixed order (say $y_0 < y_1$).\\
Alice announces $z := a\oplus b$ to \Ra\ 
where $a$ is such that $y= y_a$.}

}

The fact that this protocol is unconditionally concealing is obvious
since the commitment depends entirely on the fact that \Sa\ knows
the inverse of $y_{0}$ or $y_{1}$. Since both have a unique inverse,
it is impossible for \Ra\ to tell which one \Sa\ knows. Intuitively,
the reason why this protocol is binding is that the problem of
finding two couples $(x_0,y_0)$ and $(x_1,y_1)$ such that
$y_{0}= \pi(x_{0})$, $y_{1}= \pi(x_{1})$ and
$h_i \cdot y_0 = h_i \cdot y_1, {1\leq i< n}$ is difficult.

\proto{$NOVY/Unveil(b)$}{
\step{\Sa\/ discloses $b$ and $x$ to \Ra,}

\step{\Ra\/ checks that $y_{z\oplus b} = \pi(x)$.}
}

Naor, Ostrovsky, Venkatesen and Yung showed that it is computationally
equivalent
to cheat the  unveiling protocol or to inverse the one-way permutation.
Any
efficient algorithm to solve one problem yields an efficient algorithm
to solve the other. Their proof technique involves an algorithm
to convert any attacker $A$ to the commitment scheme to an inverter $I$
of the one-way permutation.

\subsubsection{NOVY on a quantum computer}
In order to describe the cheating protocol, we use standard tricks of
quantum computation. For the reader unfamiliar with quantum computing
we recommend \cite{andre} as an introduction. We now describe precisely
the attack.

\noindent {\bf The attack.}
In the cheating protocol, the
state $\psi = \alpha |0\rangle + \beta |1\rangle$ 
is the input bit $b$ in superposition.  
So, we denote $B$ the register that contains the state $\psi$. 

\proto{$NOVY/Commit(*)$}{
\step{\Sa\/ chooses $x$ and computes $y = \pi(x)$ at the quantum level,
that is, she sets up quantum registers $X,Y$ in state $\sumd{x\in
\BB^n}{} \frac{1}{\sqrt{2^n}} \ket{x,\pi(x)}$,}

\step{\For $i\in\{1,...,n-1\}$ \Do}

\step{\Ra\/ picks a hash vector $h_i \inr\BB^n$ and announces it to
\Sa,}

\step{Let $S_i = \{x |  h_j \cdot \pi(x) = r_j, \mbox{for} 1\leq 
j<i\}$.\\
\Sa\/ computes $r_i = h_i \cdot y$ at the quantum level, that is,
she sets up registers $X,Y,R$ in state
$\sumd{x \in S_i}{}
\frac{1}{\sqrt{2^{n-i+1}}} \ket{x,\pi(x),h_i \cdot \pi(x)}$,
and announces $r_i$, the outcome of measuring $R$, to \Ra.}

\step{\Endfor}

\{
{\rm At this point $S_{n-1}$ contains two solutions $x_0,x_1$ to
$\{ r_i = h_i \cdot \pi(x) \}_{1\leq i< n}$.\\
So the state of the registers $B,X,Y$ is 
\begin{eqnarray*}
\lefteqn{(\alpha |0\rangle + \beta |1\rangle) \otimes 
\frac{1}{\sqrt{2}} (|x_0,y_0\rangle + |x_1, y_1\rangle) }\\
&=& \frac{\alpha}{\sqrt{2}} |0,x_0,y_0\rangle
+
\frac{\alpha}{\sqrt{2}} |0,x_1,y_1\rangle
+
\frac{\beta}{\sqrt{2}} |1,x_0,y_0\rangle
+
\frac{\beta}{\sqrt{2}} |1,x_1,y_1\rangle.
\end{eqnarray*}
}
\}

\step{
\Sa\/ computes $z := a \oplus b$, where $a$ is the index of $y$,
at the quantum level, that is,
she prepares the registers $B,X,Y,Z$ in the
state 
\[
\frac{\alpha}{\sqrt{2}} |0,x_0,y_0,0\rangle
+
\frac{\alpha}{\sqrt{2}} |0,x_1,y_1,1\rangle
+
\frac{\beta}{\sqrt{2}} |1,x_0,y_0,1\rangle
+
\frac{\beta}{\sqrt{2}} |1,x_1,y_1,0\rangle.
\]
Then she measures the register $Z$ and announces $z$ to \Ra.}
}

\proto{$NOVY/Unveil(*)$}{
\step{\Sa\/ measures registers $B$ and $X$, and announces $b$ and $x$ to \Ra,}

\step{\Ra\/ checks that $y_{z \oplus b} = \pi(x)$.}

}

\paragraph{If \Sa\/  unveils.}
First we want to verify that, if \Sa\ unveils the bit,
she passes the test.  Perhaps the easier way to verify this fact
is to actually compute the state she has after the commit  part.
If she announces $z = 0$, the state of $B,X,Y$ is
\[
\alpha |0,x_0,y_0\rangle
+
\beta |1,x_1,y_1\rangle.
\]
If she announces $z = 1$, the state of $B,X,Y$ is
\[
\alpha |0,x_1,y_1\rangle
+
\beta |1,x_0,y_0\rangle.
\]
Note that in both cases, the bit $b$ has the correct distribution
of probability (the one associated with the initial state $\psi$).
Also, one may easily check that $z \oplus b = a$ so that
$y_{z\oplus b} = y = \pi(x)$.  Therefore, \Sa\ passes the test.
There is another way to see the same result. 
To unveil the bit $b$, \Sa\ measures the registers $B$ and $X$.
Because these registers are only used as control registers, 
\Sa\ could measure them at the beginning
of the protocol (just after the first step) and this would make no 
difference (as far as the distribution of probabilities is concerned).  
This can actually be verified by checking that the operation
associated with a measurement on $B$ and $X$ (in the computational
basis) and the operation associated with a computation where
$B$ and $X$ are control registers always commute. 
In the viewpoint where she measures these registers at the beginning of
the protocol, we are back to the honest classical protocol because
all superpositions disappear.  Clearly, if \Sa\  is honest she should
pass the test.

\paragraph{If \Sa\/ never Unveils.}
Second we observe that if $NOVY/Unveil(*)$ does not take place, then
\Sa\ may recover $\psi$ from her registers $B,X,Y$. 
She has only to compute $(x_a,y_a)$  at the
quantum level using $a = b \oplus z$ and then erase
the registers $X$ and $Y$ using a bitwise XOR,
and discard these registers.
Note that to compute $x_a$, she needs to compute $f^{-1}(y_a)$
because she does not know $(x_0,x_1)$, she only knows $(y_0,y_1)$.

\paragraph{Randomness without random tape.}
\Sa\ is not committed to a fixed value of $b$ in the cheating
protocol.  This is not breaking the protocol, because even
in the classical world one could easily construct \Sa's
strategy so that the attack does not define a fixed bit: she
only has to choose the bit at random.  In fact, the distribution of 
probability for the
variables in the cheating protocol is exactly the same
as in the honest protocol. So, in any reasonable
definition of security, one cannot require that \Sa\/ is committed to a fixed
bit defined by the attack. 

However, there is still a fundamental
difference between the classical situation and the quantum situation.  
In the classical world, one can look at \Sa's random tape and actually
determine the bit.  So the attack and the random tape together determine
the bit.  This is why we intuitively think that \Sa\ is committed to
a fixed bit.  This is not true anymore in the quantum case.  
We cannot think anymore that \Sa\ is committed to a fixed bit
(determined by  the value of the random tape).  
In a quantum protocol, the outcomes of measurements introduce some 
randomness which cannot be explained by the use of a random tape.  
There is no such a thing as a random tape which uniquely determine the bit.  

This means that the na\"\i ve definition of security associated with
classical bit commitment, namely that \Sa\ must be committed to a fixed bit,
is not valid anymore in the quantum world.  This is what we consider
a weak quantum attack on classical bit commitment.  This is why the security
criteria proposed in \cite{mayers97} is that \Sa\ should be committed 
to a random distribution of probability.  This notion of security
is valid in both the quantum and the classical world.

The fact that in a quantum protocol there is randomness without initial 
random tape has far reaching consequences (other than simply attacking
our na\"\i ve notion of security for classical bit commitment).  One
important ingredient in the
proof of NOVY, which reduces the security of their classical bit commitment
against \Sa\ to the existence of one way permutation, is that the randomized 
strategy used by \Sa\ is replaced by a deterministic strategy by fixing 
the value of the random tape.  Unfortunately, because randomness can
still exist in a quantum protocol even if we fix the initial random
tape, this approach does not work.  One cannot consider that \Sa\
performs a fixed strategy.  So the proof of Novy does not (at the
least not directly) apply to quantum attacks.  This is another
kind of quantum attacks against classical bit commitment: here
it's the proof of security that is attacked, not the protocol
directly.      
   
\subsection{Two-Prover Bit Commitment Scheme}
Our second example is the two-prover BC scheme of BGKW.
The assumption used for this protocol is that two parties \Sa\/ and
\Sap\/ who are allowed to exchange information before the beginning of
the protocol, cannot communicate during the execution of the protocol.
Nevertheless, both of them can talk to \Ra. This assumption may be
implemented by trapping \Sa\/ and \Sap\/ in Faraday cages or using
relativistic effects keeping them separate of a large enough distance.
In an initialization phase, \Sa\/ and \Sap\/ share information
necessary to run the commitment protocol.

\proto{$2P/Init$}{
\step{\Sa\/ picks $r \inr\BB^n$, and shares it as $r'$ with \Sap,}

\step{\Sa\/ and \Sap\/ are physically split,}
}

In order to commit they run the following
\proto{$2P/Commit(b)$}{
\step{\Ra\/ sets $m_0 := 0^n$ and picks $m_1 \inr\BB^n$, and announces
them to \Sa,}

\step{\Sa\ sends $z := r \oplus m_b$ to \Ra.}
}

The commitment is concealing because for each $z$ there exists a
unique pair $r_{0},r_{1}$ such that $r_{0} \oplus m_0 = z = r_{1} \oplus
m_1$. On the other hand, it is binding because \Sap\ does not know the
value of $m_1$.

If unveiling is required they run

\proto{$2P/Unveil(b)$}{
\step{\Sa\/ discloses $b$ and $r$ to \Ra,}

\step{\Sap\/ discloses $r'$ to \Ra,}

\step{\Ra\/ checks that $r= r'$ and that $z = r \oplus m_b$,}
}

Since \Sap\/ does not know $m_{1}$, she
is restricted to disclosing her $r'$ to have non negligeable
probability of satisfying \Ra.

Ben-Or, Goldwasser, Kilian and Wigderson \cite{BGKW} have used this protocol
to prove NP statements in {\em perfect zero-knowledge}. This follows from
the fact that this bit commitment is unconditionally concealing.

\subsubsection{Defeating $2P/Commit$}
Let $R,R'$ be a pair of quantum registers in state $(\frac{1}{\sqrt{2}}
(\ket{00}+\ket{11}) )^n$ shared between
\Sa\ and \Sap\/ before they are physically separated and let
$B:= \psi = \alpha \ket{0} + \beta \ket{1}$ be a register containing
the particle given to \Sa\/ by \Ra.

\proto{$2P/Init$}{
\step{\Sa\/ and \Sap\/ choose and share a common string at the quantum 
level, that is, they share registers $(R,R^\prime)$
in state $(\frac{1}{\sqrt{2}} \ket{00} + \frac{1}{\sqrt{2}} \ket{11} )^n$,}

\step{\Sa\/ and \Sap\/ are physically split,}

}

Commitment is performed by superposition of the honest protocol:

\proto{$2P/Commit(*)$}{
\step{\Ra\/ sets $m_0 := 0^n$ and picks $m_1 \inr\BB^n$, and announces
them to \Sa,}

\step{\Sa\ computes $z := r \oplus m_b$ at the quantum level, that is,
she prepares registers $B,R,Z$ in state
$$\sumd{r\in \BB^n}{} \frac{1}{\sqrt{2^n}}
(\alpha \ket{0,r,r \oplus m_{0}} + \beta \ket{1,r,r \oplus m_{1}})$$}

\step{\Sa\ measures $Z$ to get $z$ and sends $z$ to \Ra.}

}

After $z$ is measured the global state of registers $B,R,Z,R'$ is
$$\frac{1}{\sqrt{2}}
(\alpha \ket{0,z \oplus m_{0},z,z \oplus m_{0}} +
\beta \ket{1,z \oplus m_{1},z,z \oplus m_{1}}).$$
Unveiling is performed by measurements on both sides:

\proto{$2P/Unveil(*)$}{
\step{\Sa\/ measures $B,R_1,...,R_n$ to get
$b,r_1,...,r_n$, and
discloses $b,r$ to \Ra,}

\step{\Sap\/ measures $R_1^\prime,...,R_n^\prime$ to get
$r_1^\prime,...,r_n^\prime$ and discloses $r'$ to \Ra,}

\step{\Ra\/ checks that $r=r'$ and that $z = r \oplus m_b$,}

}

\paragraph{If \Sa-\Sap\ Unveil}
We now show that this unveiling is always successful.
Remember that after $z$ is measured, 
the global state of registers $B,R,Z,R'$ is
$$\frac{1}{\sqrt{2}}
(\alpha \ket{0,z \oplus m_{0},z,z \oplus m_{0}} + \beta \ket{1,z \oplus m_{1},z,
z \oplus m_{1}})$$ and thus
the bit $b$ has the correct distribution
of probability (the one associated with the initial state $\psi$).
Also, one may easily check that in both cases $r=z \oplus m_{b}=r'$.
Therefore, \Sa-\Sap\ pass the test.

\paragraph{If \Sa-\Sap\ never Unveil}
Second we observe that if $2P/Unveil(*)$ does not take place, then
\Sa-\Sap\ may recover $\psi$ from their registers $B,R,R'$. 
They only have to compute $z \oplus m_{b}$  at the
quantum level and then erase
the registers $R$ and $R'$ using a bitwise XOR,
and discard these registers. This computation may be done efficiently,
but it requires that \Sa-\Sap\ get back together.

\section{Discussion and Conclusions}
The first proof provided for the impossibility of bit commitment
has completely obliterated the possibility of creating an
unconditionally secure bit commitment.  However, the attack was only
indirectly described.  Subsequently, specific attempts to by-pass this
general result were proposed\cite{bc96,claude}.  This has shed more
light on the nature of the attack which was finally described
explicitly in \cite{mayers97}.  Our goal here was to provide an
analysis of this general attack in the context of a specific example,
and to create a wholeness for the different papers published on the
subject.
Moreover, we have demonstrated that it is impossible to base the security
of quantum protocols on unconditionally concealing bit commitment schemes,
even if they were proven secure in the classical world.
Notice however that it is still possible to use computationally
concealing BC protocols
such as \cite{Naor} to get a computationally secure Quantum Oblivious Transfer
\cite{crepeau94,bbcs91} protocol based on (quantum) one-way functions;
a result unlikely to be true in the classical scenario.
The big lesson to learn from all this is that quantum
information is always more elusive than its classical counterpart:
extra care must be taken when reasoning about quantum cryptographic
protocols and analyzing them.  We also hope that this paper will help to
clarify the issue of the impossibility of bit commitment in its full
generality.


\begin{thebibliography}{99}

\small

\newcommand{\andd}{{\rm and\ }}

\bibitem{dom:impos} 
{\sc Mayers,~D.}, ``Unconditionally secure
quantum bit commitment is impossible'', {\em Fourth
Workshop on Physics and Computation --- PhysComp~'96}, Boston,
November~1996.

\bibitem{mayers97} {\sc Mayers,~D.}, ``Unconditionally secure quantum
bit commitment is impossible'', {\em Physical Review Letters}, vol 78,
\mbox{pp.~3414\,--\,3417} (1997).  Note that this paper has the same
title as \cite{dom:impos} even though it uses a different approach.

\bibitem{bc96}
{\sc Brassard,~G. \andd C.~Cr\'epeau},
``Cryptology column --- 25 years of quantum cryptography'',
SIGACT News, vol. 27, no. 3, \mbox{pp. ~13\,--\,24} (1996).





\bibitem{andre} 
{\sc Berthiaume,~A.},
\newblock ``Quantum Computation.''
\newblock To appear in Complexity Theory Retrospective II,
Springer-Verlag, 1996.
\newblock Available from
``http://andre.cs.depaul.edu/andre/publications/CTR.ps.gz''.

\bibitem{claude} 
{\sc Cr\'epeau,~C.},
``What is going on with quantum bit commitment?'',
{\it Proceedings of Pragocrypt~'96: 1st International Conference on the
Theory
and Applications of Cryptology}, Prague, October~1996.



\bibitem{BGKW}
{\sc Ben-Or,~M., Goldwasser~S., Kilian,~J., and Wigderson,~A.}
``Multi-prover interactive proofs: How to remove intractability assumptions.''
In Proceedings of the Twentieth Annual ACM Symposium on Theory of Computing,
pages 113-131, Chicago, Illinois, 2-4 May 1988. 
       
\bibitem{lc97} {\sc Lo,~H.--K. \andd H.\,F.~Chau}, ``Why~quantum bit
commitment and ideal quantum coin tossing are impossible.''
Los~Alamos preprint archive {\tt quant-ph/9711065}, November~1997.

\bibitem{kent97a}{\sc Kent,~A.}, ``Quantum Bit Commitment from
a Computation Bound'', Los~Alamos preprint archive {\tt
quant-ph/9711069}, November~1997.

\bibitem{kent97} {\sc Kent,~A.}, ``Permanently secure quantum bit
commitment protocol from a temporary computation bound'', Los~Alamos
preprint archive {\tt quant-ph/9712002}, December~1997.

\bibitem{Naor}
{\sc Naor,~M.}
``Bit commitment using pseudo-randomness (extended abstract).''
In G. Brassard, editor, Advances in Cryptology --
CRYPTO '89, volume 435 of Lecture Notes in Computer Science,
pages 128-136, 20-24 August 1989. Springer-Verlag, 1990.
       
\bibitem{bb84}
{\sc Bennett,~C.\,H. \andd G.~Brassard},
``Quantum cryptography: Public key distribution and coin tossing'',
{\em Proceedings of IEEE International Conference on Computers,
Systems and Signal Processing}, Bangalore, India, \dec~1984,
\pp{175}{179}.

\bibitem{bbbss92}
{\sc Bennett,~C.\,H., F.~Bessette, G.~Brassard, L.~Salvail \andd
J.~Smolin},
``Experimental quantum cryptography'',
\jofc, Vol.~5, no.~1, 1992, \pp{3}{28}.

\bibitem{bcms97} 
{\sc Brassard,~G., Cr\'epeau,~C.,  Mayers,~D. and Salvail,~L.},
``A brief review on the impossibility of quantum bit commitment.''
Los~Alamos
preprint archive {\tt quant-ph/9712023}, December 1997. 

\bibitem{bbcs91}
{\sc Bennett,~C.\,H., G.~Brassard, C.~Cr\'epeau \andd
M.--H.~Skubiszewska},
``Practical quantum oblivious transfer'',
\crypto{91}, \pp{351}{366}.

\bibitem{bcjl93}
{\sc Brassard,~G., C.~Cr\'epeau, R.~Jozsa \andd D.~Langlois},
``A~quantum bit commitment scheme provably unbreakable by both
parties'',
\focs{34th}{\nov~1993}, \pp{362}{371}.

\bibitem{yao95}
{\sc Yao,~A.\,C.--C.},
``Security of quantum protocols against coherent measurements'',
{\it \mbox{Proceedings} of 26th \mbox{Annual} ACM Symposium
on the Theory of Computing}, 1995, \pp{67}{75}.

\bibitem{claude:louis}
{\sc Cr\'epeau,~C. \andd L.~Salvail},
``Quantum oblivious mutual identification'',
\euroc{95}{May}, \pp{133}{146}.


\bibitem{BCJL:kaput}
{\sc Mayers,~D.}, ``The~trouble with quantum
bit commitment'',
LANL Report No. quant-ph/9603015 (to be published).  The
author first discussed the result in Montr\'eal at a workshop on
quantum information theory held in October 1995.


\bibitem{lc96}
{\sc Lo,~H.--K. \andd H.\,F.~Chau},
``Is~quantum bit commitment really possible?'',
{\em Physical Review Letters}, vol 78,
\mbox{pp.~3410\,--\,3413} (1997).

\bibitem{people} {\sc Bennett,~C.\,H., C.~Fuchs, T.~Mor} (personal
communication).

\bibitem{BCC88}
{\sc Brassard,~G., D.~Chaum \andd C.~Cr\'epeau},
``Minimum Disclosure Proofs of Knowledge'',
Journal of Computer and System Sciences, vol. 37(2), 1988,
\pp{247}{268}.

\bibitem{NOVY}
{\sc Naor,~Moni, R.~Ostrovsky, R.~Venkatesan \andd M.~Yung},
``Perfect Zero-Knowledge Arguments for NP Can Be Based on General
Complexity Assumptions'', \crypto{92}, \pp{196}{214}.

\bibitem{crepeau94}
{\sc Cr{\'e}peau,~C.},
``Quantum oblivious transfer'',
{\it Journal of Modern Optics},
Vol.~41, no.~12, December~1994, \pp{2445}{2454}.


\end{thebibliography}
\end{document}